\documentstyle[12pt,aaspp4]{article}

\def\kms{\ifmmode {{\rm \;km\;s^{-1}\;}}		    	      
       \else {\hbox{$\,${\rm km$\;$s$^{\rm -1}\;$}}}\fi}
\def\solar {\ifmmode_{\mathord\odot\;} \else $_{\mathord\odot}\;$\fi} 
\def\mo {\ifmmode {\,{\it M}\solar\;} \else $\,M$\solar$\;$\fi}	      
\def\cmm#1{\ifmmode {\,{\rm cm^{-#1}}\;} 		              
	\else \hbox{$\,${\rm cm$^{\rm -#1}\;$}}\fi}
\def\am {\ifmmode {^{\scriptscriptstyle\prime}}			       
	\else $^{\scriptscriptstyle\prime}$\fi}
\def\deg {\ifmmode^\circ\else$^\circ$\fi}			       

\def\x {\ifmmode\times\else$\times$\fi}			       	       
\def\E#1 {\ifmmode {\times 10^{#1}\;} 
	\else \hbox{$\times 10^{#1}\;$}\fi} 
\def\T#1 {\ifmmode \,10^{#1}\; \else {${\rm\,10^{#1}}\;$}\fi}	       
\def\half{\ifmmode \textstyle{1\over2} \else $\textstyle{1\over2}$\fi} 

\def\apro{$\sim$}                                   
\def\masmenos{\ifmmode {\pm} \else $\pm$ \fi}		               
\def\gsim {\ifmmode {\buildrel>\over\sim}		               
	\else {\lower.6ex\hbox{$\buildrel>\over\sim$}}\fi}
\def\lsim {\ifmmode {\buildrel<\over\sim}	   	               
	\else {\lower.6ex\hbox{$\buildrel<\over\sim$}}\fi}
\def\lt{\ifmmode {<}\else{$<$}\fi}
\def\gt{\ifmmode {>}\else{$>$}\fi}
\def\tkin {$T_{\rm kin}$}

\chardef\isp="10 
\def\i{\'\isp}
\def\hdos {\hbox{${\rm H}_2$}}                    
\def\nht {\hbox{${\rm NH}_{3}$}}                  
\def\chtcn{\hbox{${\rm CH}_3{\rm CN}$}}           
\def\uc{\rm J=$1\rightarrow0\;$}					
\def\du{\rm J=$2\rightarrow1\;$}					
\def\td{\rm J=$3\rightarrow2\;$}					
\def\cc{\rm J=$5\rightarrow4\;$}					

\def\as {\ifmmode {^{\scriptscriptstyle\prime\prime}} 		       
	\else $^{\scriptscriptstyle\prime\prime}$\fi}
\def\hcop {\hbox{${\rm HCO}^+$}}                  

\slugcomment{Manuscript draft:  June 19, 1996}    
\lefthead{Mart{\i}n-Pintado \etal}
\righthead{SiO in the Galactic center}

\begin{document}

\title{SiO emission from the Galactic Center Molecular Clouds}

\author{ J. Mart{\i}n-Pintado,
P. de Vicente,  
A. Fuente,
P. Planesas}

\affil{Observatorio Astron\'omico Nacional (IGN), Campus Universitario, 
Apartado
1143, E-28800 Alcal\'a de Henares, Spain}


\begin{abstract} We have mapped  the \uc\ line of SiO in  a
1\deg\x12\am (l\x\,b) region around the Galactic center (GC) with an
angular resolution of 2\am\ (\apro4 pc). In contrast to  the spatial
distribution of other high dipole moment molecules like CS,  whose
emission is nearly uniform, the SiO emission is very fragmented and it
is only associated with some molecular clouds. In particular, it is
remarkable that the SiO emission closely follows the non-thermal radio
arc in the GC. The SiO clouds are more extended than the beam  with
typical sizes between 4 and 20 pc. High angular resolution  (26\as)
mapping in the \du line of SiO toward the molecular clouds in Sgr B2
and Sgr A shows that the SiO emission  is relatively smooth with 
structures of typically 2 pc. From the line intensities of the \du,
\td  and  \cc  transitions  of SiO we derive \hdos\ densities for
these clouds of a few  \T{4} \cmm{3}. The SiO fractional abundances
are \apro\T{-9} for the SiO clouds and \lsim\T{-10} for the other
molecular clouds in the GC. The characteristics (size and \hdos\
densities) of the SiO  emission in the GC are completely different
from those observed  in the Galactic disk, where the SiO emission
arises from  much smaller  regions with larger \hdos\ densities. We 
briefly discuss the implications of the SiO emission in the  molecular
clouds of the GC. We conclude that the particular chemistry in these
clouds is probably  related to large scale fast shocks occurring in the
Galactic center region.

\end{abstract}



\section{Introduction}

The physical conditions of the molecular clouds in the Galactic center (GC)
differ substantially from those of the molecular clouds in the Galactic disk
(see e.g. G\"usten 1989). High gas kinetic temperatures (\tkin \gsim80 K) are found
in these molecular clouds (G\"usten, Walmsley \& Pauls 1981; Morris et al. 1983;
H\"uttemeister et al. 1993). The gas kinetic temperatures are clearly above the
dust color temperatures, \apro30 K, derived from the FIR emission (Odenwald and
Fazio 1984). To explain the high kinetic temperatures of the molecular gas,
several heating mechanisms which act only on the gas have been discussed. In
particular, the large linewidth (\gsim15 \kms) that the molecular lines exhibit
in these molecular clouds and its possible correlation with the kinetic
temperature suggest that dissipation of turbulence driven by differential
Galactic rotation is an attractive heating mechanism for the GC molecular clouds
(Wilson et al. 1982; G\"usten 1989). 

If supersonic turbulence is an efficient heating mechanism, it is
expected to produce shocks which will heat the gas and also influence
the chemical composition of the molecular clouds in the GC. 
In fact, ionizing shocks     
have also been proposed as the origin of the ionization in the 
Galactic center (Morris 1989). 
Molecular species which are formed by shock chemistry are then expected to be
enhanced in these molecular clouds. Since the discovery of SiO
emission in interstellar clouds, it is rather well established that
SiO is an unambiguous tracer of high temperature and/or shock
chemistry in interstellar clouds (Downes et al. 1982; Ziurys, Fribeg 
\& Irvine 1989; Mart{\i}n-Pintado, Bachiller \& Fuente 1992 )

In this letter we report the detection of widespread SiO emission
toward some giant molecular clouds in the Galactic center. In contrast
to the star-forming regions, the SiO maps reveal very extended
emission which does not seem to be associated with recently formed
stars, but most likely with large scale shocks in the Galactic center
region.

\section{Observations and Results}

The large scale mapping of the molecular clouds in the \uc\  line of
SiO was carried out with the 14-m telescope of the Centro
Astron\'omico de Yebes (Spain). The Half Power Beamwidth (HPBW) of
the telescope was 2\am. The receiver, equipped with a cooled Schottky
mixer, had a double side band temperature of 75 K. The typical single side band          
system temperature was 250-300 K. The spectrometer
was a 512 channel acusto-optic device with a resolution of 108 kHz
(0.74 \kms). The calibration was made by using the standard chopper
wheel method. The observations were made in position switching with
the reference 15\am\ away from the Galactic plane. 
In view of the unexpected large extent of the SiO emision, the most
critical positions in our map have been checked for contamination 
with emission from the reference. 
The typical noise
in the map is 0.15 K. The SiO profiles show broad lines with typical
widths to half power of 30-60 \kms, with most of their emission
concentrated at radial velocities between $-$10 to 90 \kms. Fig. 1
shows the integrated intensity map of the \uc\ line of SiO for the
velocity range between $-$10 to 90 \kms. The SiO  emission shows a very fragmented
distribution concentrating in 17 molecular
clouds which are designated in Fig. 1 by their Galactic  coordinates.

The fragmented SiO emission is in contrast with the fairly uniform distribution
of the CS emission (Bally et al. 1987)
obtained with the same angular resolution than the SiO data. 
The  SiO emission is also different from that of CS for  negative
radial velocities. While CS shows relatively strong  emission for
radial velocities from $-$50 to $-$10 \kms toward thermal arched
filaments, SiO is not detected to our limit of 0.15 K. Along the
Galactic plane, the SiO emission shows a similar spatial distribution
to that of the hot gas observed in \nht\ emission (Morris et al.
1983). Both lines are not detected  for Galactic longitudes  between
0.3\deg\ and 0.4\deg. There are three major groups of SiO clouds. The
first one,  located south of  Sgr A, surrounds the southern edge of 
the radio continuum emission from Sgr A East. Fig. 3 shows a comparison 
of the SiO emission with the main radiocontinuum 
features within 50 pc of Sgr A*.  The second  one is found
towards the Galactic center non-thermal radio arc at l\apro0.2\deg. It
is remarkable that the SiO emission is not only restricted to the
molecular cloud M0.20-0.03 (Bally et al. 1987; Serabyn \& G\"usten
1991; Lindqvist et al. 1995), but it shows the same morphology as
the non-thermal radio arc over scales of several parsecs (see Fig. 3). The third SiO
complex is found toward the star forming regions Sgr B1 and Sgr B2. The
SiO emission does not show a clear correlation neither with the emission at 
60$\mu$m nor with the radio continuum emission (Altenhoff et al. 1979) observed 
with similar angular 
resolution. This indicates that the bulk of the
SiO clouds are not associated with newly formed OB stars. 

The SiO emission is    
more extended than the beam with typical  sizes of 4-8\am\ (9-20pc). 
High angular resolution (26\as) maps in the \du\ line of SiO of the two 
molecular clouds
associated with Sgr  A  and Sgr B2 shows that the SiO emission is not 
highly clumply in the 2\am\ beam of the \uc\ line.
These  observations were made with
the IRAM 30-m telescope at Pico  Veleta (Spain) and the observing
procedure has been described by Mart{\i}n-Pintado et al  (1992). Figs.
2a and 2b  show the integrated intensity maps of the
\du\  line of SiO towards the Sgr B2 and Sgr A  molecular clouds
respectively. These data will be analyzed in more detail elsewhere
(Mart{\i}n-Pintado et al 1996). Simultaneously with the \du\ line we
also  observed the \td\ and  \cc\ lines of SiO. 
The HPBW of the telecopes for the \td\ and  \cc\ lines were 17\as\ and 13\as\
respectively. To derive the physical conditions, 
fully sampled maps (5\x5 point) in the \cc\ line were also made at selected positions. 
Fig. 4 shows the
typical line profiles of the SiO lines towards two positions in the
Sgr B2 molecular cloud.  The SiO profiles are very broad with
linewidths to zero  intensity up to 100 \kms.


The SiO  emission south of Sgr A (Fig. 2b) is elongated (8 pc\x16 pc)
along  the Galactic plane with two main clouds, M-0.13-0.08 (the 20
\kms cloud) and M-0.02-0.07 (the 50 \kms  cloud) and a  condensation
close to Sgr A*. Like at large scale, the overall morphology of the 
SiO emission from these clouds is similar to that of the hot gas
observed in the (3,3) line of \nht (G\"usten, Walmsley \& Pauls 1981;
Ho et al. 1991). The SiO emission, however,  does not show the
prominent cold and dense  FIR condensations observed at 1.3 mm (Mezger
et al. 1989). The SiO  emission toward Sgr B2 extends over a region of
at least \apro24\x24 pc and presents a fragmented  distribution. The
typical  size of the SiO condensations is \gsim1\am (\gsim2 pc).
The SiO lines show doubled peaked profiles. While the gas with low radial
velocities, \apro0\kms, appears to the east of the star forming region, the
high velocity gas, 60-80\kms, appears mainly to the west. The very broad
SiO lines and the systematic trend observed in radial velocities suggest that in the
envelope of Sgr B2  very energetic events are taking place. Furthermore, from
the integrated emission one can recognize several
shell-like structures. The largest one, centered on Sgr B2M,
surrounds the hot ring observed in \chtcn ( de Vicente et al. 1996)

\section{Physical conditions in the SiO clouds}

The main characteristic of the SiO lines in the GC  clouds is  that
the intensity of the \cc\ line is typically  6-10 times weaker than
that of the \du line, except for the star-forming regions Sgr B2M and
Sgr B2N. 
The line intensity ratios have been derived after smoothing the \cc and the \td lines to
the the same resolution of the \du line.
The low intensity ratios in the GC clouds is in contrast with  those found in
star-forming regions in the Galactic disk where ratios  of \apro1 are
observed (Mart{\i}n-Pintado, Bachiller \& Fuente 1992). The low 
\cc/\du line intensity ratio  in the GC suggests that the SiO emission
in these clouds arises from  material with moderate \hdos\ densities.
Model calculations of  the excitation of SiO using the Large Velocity
Gradient  Approximation (LVG) indicate that \hdos\ densities of a few \T{4}
\cmm{3} can explain the observed line intensity ratios for  the
typical kinetic temperature, \gsim50 K, of the GC  molecular clouds
(G\"usten, Walmsley \& Pauls 1981; Morris et al. 1983; H\"uttermeister
et al. 1993). The derived \hdos\ densities are basically independent of
the kinetic temperatures for kinetic temperatures larger than 50 K. 
For the typical physical conditions of the GC molecular clouds  (\hdos\  density
of 5\T{4} \cmm{3} and a kinetic temperature of 80 K), the LVG analysis  
gives SiO column densities of 0.7-3\T{14} \cmm{2}.

Asuming the same physical conditions for the clouds  where the SiO emission is not
detected, we derive upper limits  to the SiO column density of
\apro\T{13} \cmm{2}. For both types  of clouds, the column densities
estimated from LVG calculations for CS and  \hcop\ are  typically of 2-5\T{14}
\cmm{2},
similar to those derived from
multitransition LVG analysis of CS towards some molecular clouds in the
GC (Serabyn \&  G\"usten 1987). The 
derived SiO column densities are similar to those of CS for the SiO clouds and a factor
of, at least, 10 smaller for the other clouds. Since SiO and CS  have
similar dipole moments and energy level distributions, the derived column density
ratio between these molecules is basically independent of the assumed physical
conditions. Assuming the standard fractional abundance for CS and \hcop, we
estimate the  fractional abundance of SiO for the SiO clouds to be  \apro\T{-9}.
The fractional abundance of  SiO is,
at least, one order of magnitude smaller, \lsim\T{-10} , for the
clouds where SiO emission has not been  detected. Similar upper limits
to the  SiO abundances are derived for the  molecular material with
negative radial velocities  associated with the thermal arched
filaments  in the GC. From the derived \hdos\ densities and the
measured sizes, the SiO clouds have typical masses of a few\T{5} \mo.

\section{Discussion}

In the molecular clouds of the Galactic disk, the SiO emission is 
mainly associated with energetic  mass outflows powered by young
stars. This peculiar chemistry has led to the conclusion that Si is
highly depleted in the molecular clouds and SiO appears only in very
small regions where shock disruption of grains releases Si or SiO to
the gas phase  (Mart{\i}n-Pintado, Bachiller \& Fuente 1992).  The
variation of the SiO abundance in the GC  molecular clouds also
indicates a peculiar chemistry for this molecule in the GC region. 
However, the main  characteristics (very widespread and moderate
\hdos\ densities) of the SiO emission in the GC are  substantially
different from those observed in the Galactic  disk. The GC SiO
molecular clouds seem to be associated with warm gas, and high 
temperature chemistry (Ziurys, Fribeg \& Irvine 1989; Langer \&
Glassgold 1990) could explain the SiO  abundances if Si is less
depleted in the GC clouds than in the clouds of the disk.  Also,
desorption of silicon bearing compounds from warm grains
could explain the large SiO abundance in the GC clouds (Turner 1992a;
MacKay 1995). However,  chemistry schemes  based only on high
temperatures cannot explain the low SiO emission abundance found in 
the hot (\gsim80 K, Serabyn \&  G\"usten 1987) material
associated with  the thermal arched filaments. The low SiO abundance
in the  thermal arched filaments is very likely related to the heating
of these filaments. The main heating mechanism is thought to be the 
UV radiation from OB stars (Poglitsch et al. 1991). This  will produce
hot photodissociation regions but low SiO abundances. The low SiO/CS abundance 
ratio is unlikely to be related to the photodissociation of SiO since this
molecule is expected to
be more resistent than CS (Sternberg \& Dalgarno 1986). 

It has been proposed that the 
high temperatures in most of the GC
clouds are related to cloud-cloud collisions which are expected to be
more frequent in the GC region than in the disk (Wilson et al. 1982).
Low velocity shocks associated with cloud-cloud collisions have also
been claimed to explain the SiO abundances derived from absorption
lines toward Sgr B2M (H\"uttemeister et al. 1995; Peng et al. 1996).
The presence of low-velocity shocks were inferred from the narrow 
linewidths (\apro10 \kms) of the absorption lines. However, the
absorption lines only sample a very particular line of sight of the
envelope and the line width might not represent the complete
kinematics of the envelope. Indeed,  the SiO emission in the envelope
of Sgr B2 (Fig. 2) shows much broader profiles than the absorption
lines with radial velocities of up to \masmenos50 \kms. For shock
velocities larger than 40 \kms grain destruction becomes important
(Seab \& Shull 1983; Tielens et al. 1994) and Si and/or SiO can be
released to gas phase. Therefore the SiO emission in the GC molecular
clouds could be associated with  relatively fast shocks.

The fast shocks in the SiO  molecular clouds are very likely of
different origin for the different SiO complexes. For the molecular
clouds south of Sgr A, there are several evidences suggesting that
these molecular clouds are  interacting with nearby supernova remnants
(Ho et al. 1991; Mezger et al. 1989). In fact the broad lines and the
large SiO abundances in the GC  clouds resemble those observed in the
molecular gas interacting with the supernova remnant IC443 (Turner et
al. 1992b). The clouds SiO+0.17-0.01 and SiO+0.20-0.07 clearly follow
at large scale the GC radio arc suggesting a physical association with
this feature (Fig. 3). The SiO emission probably traces the interaction of the
molecular material with the non-thermal filaments. Strong shocks will
occur either in the case that the relativistic particles present along
the filaments impact on the molecular cloud or, conversely, that the
relativistic particles originate in the molecular cloud by magnetic
field reconnection between the magnetic fields in the molecular cloud
and in the radio arc  (Serabyn \& G\"usten 1991; Serabyn \& Morris 1994).
The origin of strong shocks in the SiO clouds of the Sgr B complex
could be due to large scale cloud collisions (Hasegawa et al 1994),
expanding bubbles driven by supernovas or  HII regions (Soufe 1990, de
Vicente et al. 1996) and sources with strong stellar winds like Wolf
Rayet stars (Mart{\i}n Pintado et al. 1996). Further high angular
resolution observations of  molecular and atomic lines are needed to
establish the origin of the peculiar chemistry in the GC molecular
clouds.

\acknowledgments We would like to thank the staffs of the 14-m and
30-m telescopes for support during the observations and P. T. P. Ho for 
the critical reading of the manuscript. This work has 
been partially supported by the Spanish CICYT under grant  number
PB93-048.



\begin{plate} 
\figurenum{1} 
\caption{{} 
Fig. 1.-- Integrated line intensity ($-$10 to 90 \kms) map of the \uc\ SiO
line  toward the Galactic center. The beam size is shown as a open
circle in the  upper right corner. The dots show the positions where
the spectra were taken.  The contour levels are 7.3 (4$\sigma$) to 94.5 by 17.4 K
\kms. 
Fig. 2.-- a) Integrated line intensity map of the \du\ SiO line toward the
Sgr B2  molecular cloud. The beam size is shown as a open circle in
the lower right corner and the dots show the positions where the
spectra were taken. The filled star shows the position of Sgr B2M.
The dashed contours correspond to absorption lines
observed toward the continuum sources Sgr B2M and Sgr B2N. The contour
levels are: -2, 9.2 18.5, 27.7, 36.9 and 41.5 K \kms.
b) Integrated line intensity map of the \du\ SiO line toward the Sgr A
molecular  clouds. The beam size is shown as an open circle in the
lower left corner. The open star shows the location of Sgr ${\rm A^*}$
and the filled squares the FIR  sources observed in the continuum
emission at 1.3mm (Mezger et al. 1989). The  dots show the
positions where the spectra were taken. The contour levels are: 16 to 105 by 5 
K \kms. The location of the two major clouds in the map is shown.}
\end{plate}

\newpage

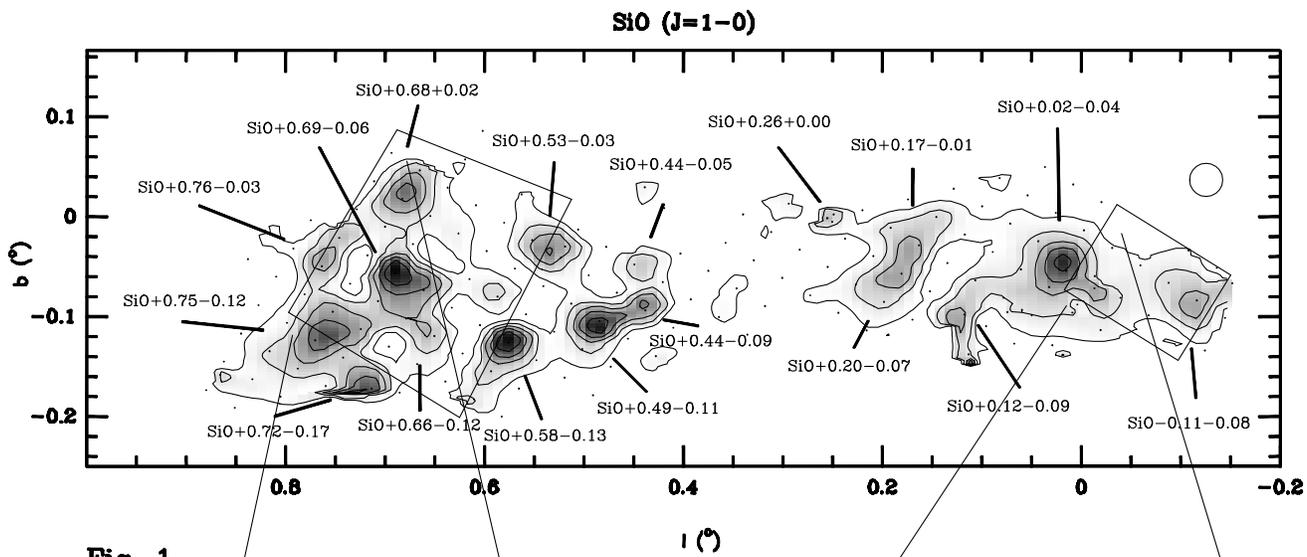
\begin{figure} 
\figurenum{3} 
\caption{{} 
The spatial distribution of the \uc\ SiO emission (thin contours) towards 
the Galactic Center Arc superimposed on that of the 20 cm radio continuum 
emission outlined by doubled thick contours (Yusef-Zadeh 1986). 
The position of SgrA* is shown by a star. The strong SiO emission is in 
general anticorrelated with the most outstanding features in radio continuum 
emission like the surroundings of Sgr A and the thermal filaments. There is 
however an outstanding coincidence of the SiO emission with the non-thermal Arc. 
This is very likely related to the origin of SiO in the galactic Center (Section 4).}
\end{figure}

\begin{figure} 
\figurenum{4} 
\caption{{} 
Sample of the \du, \td\ and \cc SiO line profiles taken toward two
positions  in the Sgr B2 molecular cloud. The offsets in the upper
right corner are in arcseconds and refer to the position of Sgr B2M. The line
intensity is in units of antenna temperature.}

\end{figure}

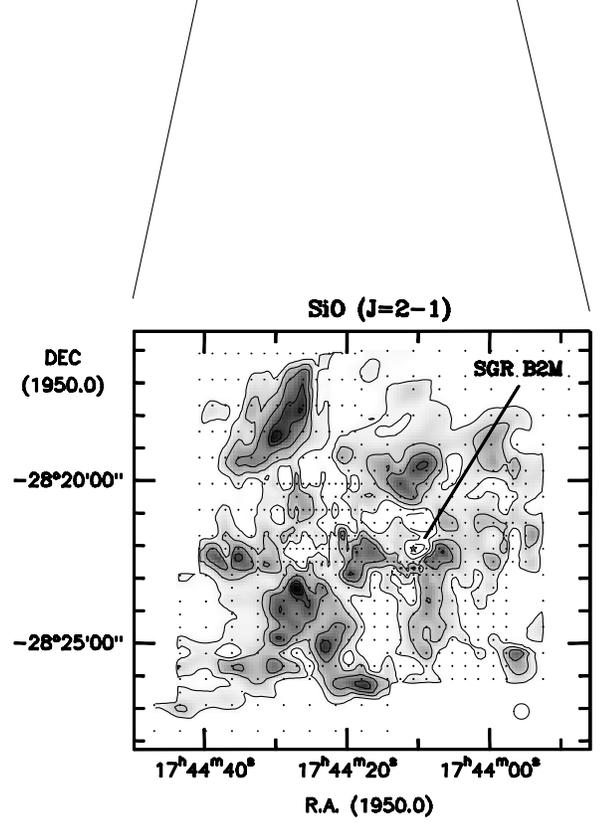
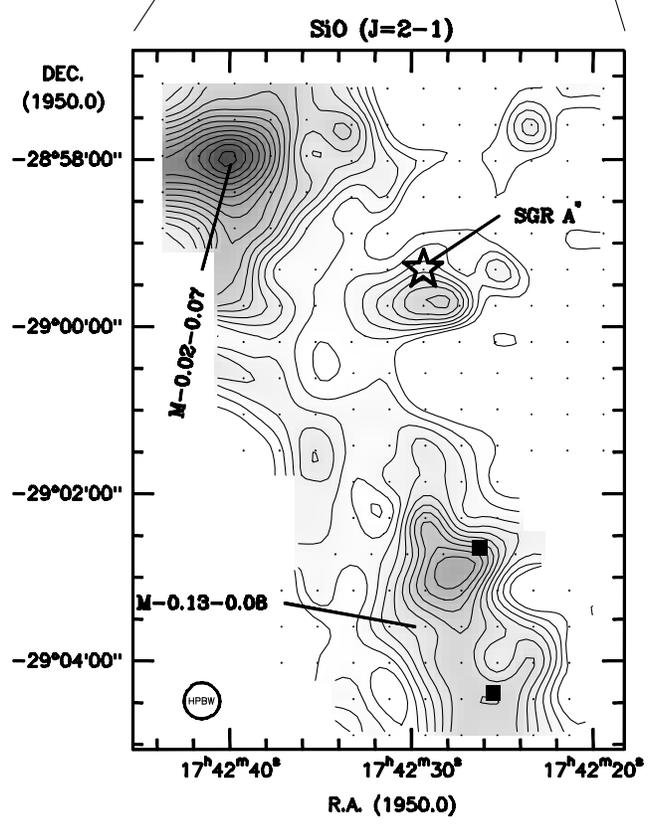
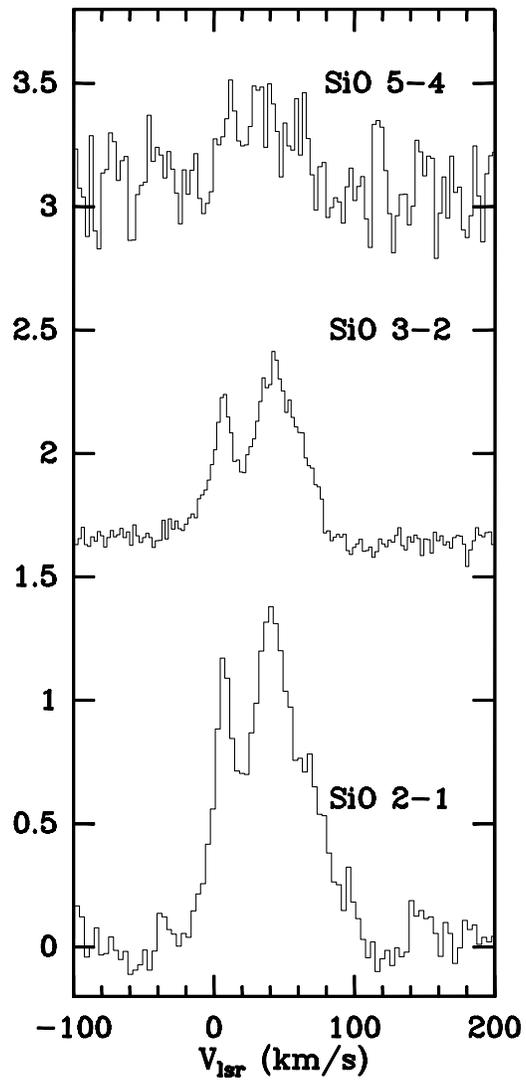
\end{document}